\documentclass[12pt,twocolumn]{aastex6}
\newcommand{\isotope}[2]{${}^{#1}$#2}
\newcommand{\msun}{\mbox{$\mathrm{M_{\odot}}$}}
\newcommand{\mdot}{\mbox{$\mathrm{\dot{M}}$}}
\newcommand{\pyr}{\mbox {{\rm yr$^{-1}$}}}
\newcommand{\mwd}{\mbox {{\rm M$_{\rm WD}$}}}
\newcommand{\gcc}{\mbox {{\rm g~cm$^{-3}$}}}

\shorttitle{Type Ia SNe keep memory of their progenitor metallicity}
\shortauthors{Piersanti et al.}

\begin{document}

\title{Type Ia Supernovae keep memory of their progenitor metallicity}

\author{Luciano Piersanti\altaffilmark{1,2}, Eduardo Bravo\altaffilmark{3}, Sergio Cristallo\altaffilmark{1,2}, Inmaculada Dom\'\i nguez\altaffilmark{4},
Oscar Straniero\altaffilmark{1,5}, Amedeo Tornamb\'e\altaffilmark{6}, Gabriel Mart\'\i nez-Pinedo\altaffilmark{7,8}}
\affil{\altaffilmark{1}INAF-Osservatorio Astronomico di Teramo, via Mentore Maggini, snc, I-64100, Teramo, Italy}
\affil{\altaffilmark{2}INFN-Sezione di Perugia, via Pascoli IT}
\affil{\altaffilmark{3}E.T.S. Arquitectura del Vall\'es, Universitat Polit\`ecnica de Catalunya, Carrer Pere Serra  1-15, 08173 Sant Cugat del Vall\`es,Spain,}
\affil{\altaffilmark{4}Universidad de Granada, E-18071 Granada, Spain}
\affil{\altaffilmark{5}INFN, Laboratori Nazionali del Gran Sasso (LNGS), 67100 Assergi, Italy}
\affil{\altaffilmark{6}INAF-Osservatorio Astronomico di Roma, via Frascati, 33, I-00040, Monte Porzio Catone, Italy}
\affil{\altaffilmark{7}GSI Helmholtzzentrum f{\"u}r Schwerioneneforschung, Planckstra{\ss}e 1, 64291 Darmstadt, Germany}
\affil{\altaffilmark{8}Institut f{\"u}r Kernphysik (Theoriezentrum), Technische Universit{\"a}t Darmstadt, Schlossgartenstra{\ss}e 2, 64289 Darmstadt, Germany}

\begin{abstract}
The ultimate understanding of Type Ia Supernovae diversity is one of the most urgent issues to 
exploit thermonuclear explosions of accreted White Dwarfs (WDs) as cosmological yardsticks. 
In particular, we investigate the impact of the progenitor system metallicity on the physical and 
chemical properties of the WD at the explosion epoch. 
We analyze the evolution of CO WDs through the accretion and simmering phases by using 
evolutionary models based on time-dependent convective mixing and an extended nuclear network 
including the most important electron captures, beta decays and URCA processes. 
We find that, due to URCA processes and electron-captures, the neutron excess and density 
at which the thermal runaway occurs are substantially larger than previously claimed. Moreover, we find 
that the higher the progenitor metallicity, the larger the neutron excess variation during the accretion and 
simmering phases and the higher the central 
density and the convective velocity at the explosion. Hence,  the simmering phase acts as an amplifier of the 
differences existing in SNe Ia progenitors. 
When applying our results to the neutron excess estimated for the Tycho and Kepler young Supernova 
remnants, we derive that the metallicity of the progenitors should be in the range ${\rm Z=0.030-0.032}$, close 
to the average metallicity value of the thin disk of the Milky Way. 
As the amount of \isotope{56}{Ni} produced in the explosion depends on the neutron excess and central density at the
thermal runaway, our results suggest that the light curve properties depend on the progenitor metallicity. 
\end{abstract}


\section{Introduction} \label{s:intro}

Nowadays it is largely accepted that Supernovae Ia (SNe Ia) are produced by the thermonuclear disruption of CO 
white dwarfs (WDs) accreting matter from their companions in binary systems \citep{hoyle1960}. In fact, the growth 
in mass of the degenerate object determines its compressional heating. For intermediate 
values of the accretion rate ($10^{-8}\le\mdot\le 10^{-7}$\msun\ \pyr)
when the accreting WD approaches the Chandrasekhar mass limit, C-ignition occurs 
at the center where the effects of compressional heating are larger \citep[e.g., see][]{piersanti2003a}.  
The energy delivered by C-burning increases the local temperature and, hence, the burning rate speeds up. 
The large energy excess can not be transferred radiatively, so the WD inner zones become unstable 
for convection. As C-burning proceeds, the convective zone extends outwards in mass, while temperature in the 
inner zones continues to increase, until it attains ${\rm \sim 8\times 10^{8}K}$ and the explosion 
occurs \citep{wunsch2004}. The time from the onset 
of convection to explosion lasts several $10^4$ yr and it is usually called {\sl simmering phase}.

The accretion and simmering phases determine the WD properties 
at the explosion. In particular, the energy and chemical composition of the ejecta are determined by the 
thermonuclear flame propagation whose properties depend on the thermal and turbulent state of the WD 
\citep{nomoto1984,woosley1986}. Moreover, the dimension and distribution of ignition points at the 
explosion depend on the temperature fluctuations already present in the convective core of the exploding WD 
\citep{garcia1995,wunsch2004}. Finally, the pre-explosive evolution fixes the WD neutron excess $\eta$, 
thus determining the amount of \isotope{56}{Ni} produced during the explosion and, hence, the 
luminosity at maximum of SNe Ia \citep{timmes2003,bravo2010}. All these effects are thought to contribute 
to the SNe Ia homogeneity, by erasing the identity hallmarks of the exploding WD progenitor 
\citep[the so-called ``stellar amnesia'' - ][]{hoflich2003}. However, such an amnesia cannot be global, as suggested 
by the correlation observed between SNe Ia luminosity and metallicity of their environment  
\citep[e.g.][]{moreno2016}. 

As mass is transferred, the WD contracts and density increases. When the electron Fermi energy 
becomes large enough, electron-capture processes occur. The capture threshold is typically lower for odd-mass 
nuclei due to odd-even effects in the nuclear masses. Electron-capture is followed by the beta-decay of the 
daughter nucleus producing an URCA process, in which electron-capture on the isobar with charge $Z$ alternates  
with the beta-decay of the $(Z-1)$ isobar. During this process, the abundances are not changed, but due to the 
$\nu_e$ and $\bar{\nu}_e$ emission substantial cooling occurs~\citep{tsuruta1970}. The cooling efficiency is 
larger at the so-called URCA density, $\rho_{\text{URCA}}$ for which electron-capture and beta-decay processes 
operate with the same rate. Increasing density, electron-capture on even-mass nuclei becomes possible. Due 
to pairing effects, the capture threshold for a second electron-capture on the produced odd-odd nucleus is much 
lower than for the first capture in the even-even nucleus. Hence, there is no URCA process but a double 
electron-capture that heats as the second capture populates final states at high excitation energy decaying by gamma 
emission. We denote $\rho_{\text{2EC}}$ the density at which the electron Fermi energy becomes equal to 
the threshold for the first capture. In stellar environment URCA processes in odd-mass nuclei and 
double electron-capture in even-mass nuclei operate simultaneously. Typically, URCA processes dominate the 
temperature evolution, as they operate for several cycles, and double electron-captures the 
neutronization, as even-mass nuclei are more abundant.

\citet{paczynski1972} early recognized that URCA processes could play an important role during the 
simmering phase. In the WD regions where the density is close to $\rho_{\text{URCA}}$, the so-called 
URCA shell, both nuclides participating in URCA processes have similar abundances. 
The continuous density increase in the accreting WD causes the URCA shell to move outwards in mass, 
and the region interior to the URCA shell becomes enriched in the proton-poor isobar, $^{A}(Z-1)$, i.e. it 
becomes neutronized. Similarly, for even A nuclei 
the transition from regions enriched in the $^{A}(Z-2)$ isobar to those enriched in the $^{A}Z$ isobar 
occurs at densities close to $\rho_{\text{2EC}}$. During the accretion phase URCA processes and double 
electron-captures operate locally. After C-ignition, 
when convection sets in, convective URCA and double electron-captures followed by double-beta decays occur. 
If the proton-poor isobar of an URCA pair or double electron-capture triplet is transported by 
convective eddies across the URCA shell to lower densities, it beta-decays to the proton-rich isobar. 
Afterward, the convective eddies move the latter once more to the high-density regions of the WD and 
electron-capture occurs. In this process, the global abundances are not changed and there is no net 
neutronization, but, due to $\nu_{e}$ and $\bar{\nu}_{e}$ emission, a substantial cooling occur. During 
the explosion, the inner $\sim 0.2$\msun~ of the WD are incinerated, and neutronized to values in excess 
of those found at the end of the simmering phase.

In the past many works were devoted to explore the effect of URCA processes during the simmering phase 
\citep{bruenn1973,couch1975,iben1978a,iben1978b,iben1982,barkat1990,mochkovitch1996,stein1999,kogan2001,lesaffre2005}. 
\citet{denissenkov2015} included URCA processes in the evolution of accreting CO WDs and showed that they 
affects the physical properties at the C-ignition epoch and, hence, the density at the explosion.
\citet{piro2008} derived $\eta$  at the explosion using a semi-analytical model 
for the \isotope{12}{C} consumption during the simmering phase. 
They did not address the effects of URCA processes and derived the neutronization independently of the progenitor metallicity.
A similar evaluation was performed by \citet{chamulak2008} by means of one-zone model coupled to a nuclear network. 
More recently, \citet{martinez2016} performed the first consistent evaluation of the neutron excess at the explosion epoch. 
They computed evolutionary model of WDs from the accretion phase up to the explosion by considering explicitly the 
\isotope{23}{(Na,Ne)} and \isotope{25}{(Mg,Na)} URCA pairs. 
The final neutronization they derived is dominated by convective URCA processes and only mildly dependent on the 
progenitor metallicity.
All these works obtained a small increase of the central neutronization during the simmering phase 
($0.3\times 10^{-3} - 10^{-3}$), definitively too low to explain the neutron excess in the exploding WD 
as estimated for the Supernova remnants in the thin disk of the Galaxy \citep{badenes2008,park2013}.

In this work we investigate the dependence of the physical properties and neutronization at the explosion 
on the accretion and simmering phases. To this aim we consider WD progenitors with different initial metallicities 
and we adopt an extended nuclear network. 
In Sect. \ref{s:models} we present the initial CO WD models and illustrate 
the computational method; in Sect. \ref{s:results} we present our results; in 
Sect. \ref{s:conclu} we summarize our conclusions.

\section{Initial models and Assumptions}\label{s:models}

We assume the Double Degenerate scenario for SNe Ia progenitors \citep{iben1984} and we 
consider three different initial binary systems, namely $\mathrm{(M_1,M_2)=}$ (3.2,1.5), (4.5,1.8) and (5.3,2.0) 
\msun\, having initial chemical composition $\mathrm{(Y,Z)=}$ (0.245,2.45$\times 10^{-4}$) (model ZLOW), 
(0.269,1.38$\times 10^{-2}$)  (model ZSUN) and (0.305,4$\times 10^{-2}$)  (model ZHIG), respectively. 
Heavy elements in ZSUN and ZHIG models have a scaled-solar abundances, while for the ZLOW model we
assume $\text{[Fe/H]}=-2$ and  $[\alpha/\text{H}]=0.5$\footnote{We adopt the heavy elements distribution provided by 
\citet{lodders2003} so that $Z_\odot=1.38\times 10^{-2}$ \citep[see][]{piersanti2007}.}.
Computations have been performed with the FUNS code \citep{straniero2006,cristallo2009}, from 
the pre-main sequence phase up to the tip of the AGB. 
The primary component of each binary system evolves into a 
CO WD with total mass \mwd=0.817 \msun, while the secondary evolves into a $\mathrm{\sim 0.6}$
\msun\, CO WD, the exact mass value depending on the considered case. 
The more massive WDs are evolved along the cooling sequence for $\mathrm{\simeq 1}$ Gyr, until 
the models attain the following physical conditions: $\mathrm{L_{sup}=1.76\times 10^{-3}L_\odot}$, 
$\mathrm{T_{eff}=1.20\times 10^{4} K}$, $\mathrm{T_C=1.1\times 10^{7} K}$ and $\mathrm{\rho_C=1.22\times 
10^{7}}$ \gcc. Due to gravitational wave radiation, 
the secondary WD undergoes a dynamical mass transfer via Roche lobe 
overflow, disrupting in a massive accretion disk around the primary WD. Hence, we assume that matter is 
transferred from the disk at \mdot=$10^{-7}$\msun\pyr. The chemical composition of the accreted matter is 
fixed by assuming that during the dynamical mass transfer the material of the original secondary WD is fully mixed. 
\begin{deluxetable}{l|cr|l}
\centering
\tablecaption{URCA pairs (lines 1-8) and double electron-capture
  triplets (lines 9-10) considered in the present work.\label{tab1}}
\tablehead{
\colhead{Isobars} & \colhead{$\rho_{\text{URCA}}$ or $\rho_{\text{2EC}}$} &
 \colhead{$\mathrm{X_\odot}$\tablenotemark{b}} & \colhead{Source} \\
\colhead{ } & \colhead{in $10^9$ \gcc} &
 \colhead{} & \colhead{} 
}
\startdata
		\isotope{19}{F} - \isotope{19}{O} & 2.43 & 1.07$\times 10^{-7}$ & Suzu2016 \\
		\isotope{21}{Ne} - \isotope{21}{F} & 3.78 & 3.74$\times 10^{-5}$ & Suzu2016 \\
		\isotope{23}{Na} - \isotope{23}{Ne} & 1.86 & 1.42$\times 10^{-4}$ & Suzu2016 \\
		\isotope{25}{Mg} - \isotope{25}{Na} & 1.31 & 3.84$\times 10^{-5}$ & Suzu2016 \\
		\isotope{27}{Al} - \isotope{27}{Mg} & 0.104 & 5.60$\times 10^{-5}$ & Suzu2016 \\
		\isotope{31}{P} - \isotope{31}{Si} & 1.09 & 6.68$\times 10^{-6}$& Oda1994 \\
		\isotope{37}{Cl} - \isotope{37}{S} & 2.19 & 3.03$\times 10^{-6}$& Oda1994 \\
		\isotope{39}{K} - \isotope{39}{Ar} & 0.012 & 3.39$\times 10^{-6}$& Oda1994 \\
\hline
		\isotope{32}{S} - \isotope{32}{P} - \isotope{32}{Si} & 0.144 & 3.14$\times 10^{-4}$& Oda1994 \\
		\isotope{56}{Fe} - \isotope{56}{Mn} - \isotope{56}{Cr} & 1.27 & 1.05$\times 10^{-3}$& Lang2001\\
\enddata
\tablenotetext{b}{Mass fraction abundance of the $\beta$-stable isotope in the initial ZSUN model.}
\tablecomments{Suzu2016 : \citet{suzuki2016}; Oda1994 : \citet{oda1994}; Lang2001: \citet{langanke2001}. 
}
\end{deluxetable}

To describe properly the evolution of accreted WDs through the simmering phase and up to the explosion, 
the FUN code solves simultaneously the equations describing the evolution of 
the stellar physical structure and those describing the evolution of chemicals, as determined by both nuclear 
burning and convective mixing. In convective zones we model the mixing as 
an advective process, adopting as advection velocity the values computed according to the mixing length theory 
of convection \citep{cox68}. In our computations 
we use a large nuclear network including 300 processes and 81 isotopes, namely: $n, p,  \alpha$, C (12-14), N (13-15), 
O (16-19), F(18-21), Ne (20,23), Na (22-25), Mg (23,27), Al (25-28), Si (27-32), P (30-33), S (31-37), Cl (35-38), 
Ar (36-42), K (38-43), Ca (40-46), SC (45-46), \isotope{56}{Cr}, \isotope{56}{Mn} and \isotope{56}{Fe}. 
The URCA processes we consider are listed in Table \ref{tab1}. 
Rates provided by \citet{oda1994} are interpolated by adopting the \citet{fuller1985} effective formalism. 
The rates for \isotope{56}{Fe} - \isotope{56}{Mn} - \isotope{56}{Cr} have been recomputed in a fine density and
temperature grid based on the same input used in~\citet{langanke2001}.
The $^{13}\text{N}(e^-,\nu_e)^{13}$C rate is computed using the experimental data provided by \citet{zegers2012}.
The \isotope{12}{C}+\isotope{12}{C} reaction rate is taken from \citet{caughlan1988}.
We assume that the explosion of the accreted WD occurs when the maximum temperature 
attains $T_{\text{exp}}=8\times 10^{8}$~K, i.e. 
when the nuclear heating timescale approaches the turnover timescale of the convective region \citep{wunsch2004}.

\section{Results}\label{s:results}

\begin{figure}[t]
   \includegraphics[width=\columnwidth]{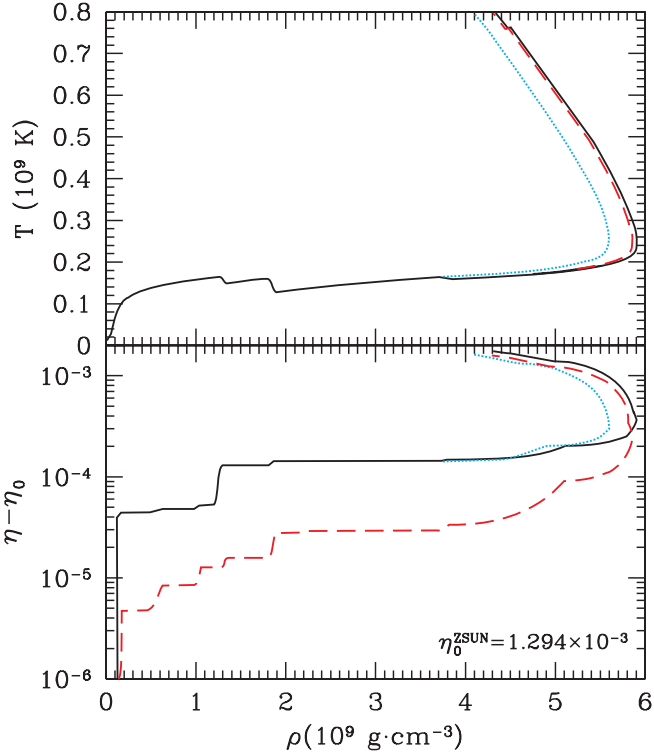}
    \caption{Upper panel: evolution in the $\rho-T$ plane of the center of the model ZSUN (solid line). 
             Lower panel: evolution of the neutronization variation at the center 
               $\Delta\eta=\eta(t)-\eta_0$ as a function of the central density (solid line). 
               Dashed and dotted lines refer to the NoSFe and NoNe21 models, respectively. 
    \label{fig1}}
\end{figure}

The compression of the accreting WD via mass deposition determines the homologous increase of density, 
while the evolution of the temperature profile is determined by the exact value adopted for \mdot\, 
\citep[e.g., see ][]{piersanti2003a}. For intermediate values of \mdot, as the one adopted in this work, 
initially the gravitational energy released on the surface of the accreting WD causes the local increase of temperature. 
When the evolutionary time becomes comparable to the thermal diffusion timescale, the temperature at the 
center increases. Later on and up to the C-ignition epoch, the temperature evolution is determined by the 
homologous compression. 
This is clearly illustrated in the upper panel of Fig.~\ref{fig1}, where we report the evolution of the center of the ZSUN 
model (solid line).

\begin{deluxetable}{lrrr|rr}
\centering
\tablecaption{Model inputs and results. \label{tab2}}
\tablehead{
\colhead{Model} & \colhead{ZLOW} & \colhead{ZSUN} & \colhead{ZHIG} & \colhead{NoSFe} & \colhead{NoNe21}
}
\startdata
$\mathrm{M_{MS}}$  (\msun)                 & 3.2   & 4.5   & 5.3   & 4.5   & 4.5   \\ 
$\mathrm{Z_{ini}}$ ($\mathrm{10^{-3}}$)    & 0.245 & 13.80 & 40.00 & 13.80 & 13.80 \\ 
$\mathrm{M_{acc}}$ (\msun)                 & 0.566 & 0.567 & 0.564 & 0.567 & 0.566 \\ 
$\mathrm{t_{acc}}$ (10$^6$ yr)             & 5.663 & 5.671 & 5.639 & 5.669 & 5.662 \\ 
$\mathrm{t_{simm}}$ (10$^4$ yr)            & 2.872 & 2.822 & 3.461 & 2.794 & 2.846 \\ 
$\mathrm{\rho_{ign}}$ ($10^9$\gcc)         & 3.727 & 4.509 & 5.305 & 4.519 & 4.350 \\
$\mathrm{T_{ign}}$ ($10^8$ K)              & 1.947 & 1.671 & 1.494 & 1.683 & 1.729 \\
$\mathrm{\rho_{simm}}$ ($10^9$\gcc)        & 4.042 & 5.113 & 6.250 & 5.085 & 4.883 \\
$\mathrm{T_{simm}}$ ($10^8$ K)             & 2.023 & 1.778 & 1.624 & 1.786 & 1.831 \\
M(\isotope{12}{C})$_{ini}$ (\msun)         & 0.562 & 0.548 & 0.503 & 0.548 & 0.548 \\
M(\isotope{12}{C})$_{fin}$ (\msun)         & 0.536 & 0.521 & 0.468 & 0.521 & 0.510 \\
$\Delta$M(\isotope{12}{C})($10^{-2}$\msun) & 2.620 & 2.763 & 3.585 & 2.732 & 2.779 \\
$\mathrm{M^{max}_{conv}}$ (\msun)          & 1.267 & 1.239 & 1.217 & 1.235 & 1.238 \\
$\mathrm{M_{exp}}$ (10$^{-2}$ \msun)       & 0.000 & 0.000 & 1.486 & 0.000 & 0.000 \\
$\mathrm{\rho_{exp}}$ ($10^9$\gcc)         & 3.374 & 4.299 & 5.073 & 4.287 & 4.068 \\
$\mathrm{\eta_{exp} (10^{-3})}$            & 1.149 & 2.996 & 6.478 & 2.884 & 2.924 \\
$\mathrm{v_{exp}}$ (km/s)                 &   554 &   713 &   863 &   710 &   672 \\
$\mathrm{\overline{\eta}_{exp} (10^{-3})}$ & 0.668 & 2.368 & 5.472 & 2.262 & 2.308 \\
$\mathrm{\eta_{c,0} (10^{-3})}$            & 0.022 & 1.294 & 3.752 & 1.294 & 1.294 \\
\hline                                                                              
\enddata
\tablecomments{${\rm M_{MS}}$: Progenitor mass in the main sequence phase; 
${\rm M_{acc}}$: total accreted mass; ${\rm t_{acc}}$: time from the onset 
of mass transfer up to the explosion; ${\rm t_{simm}}$: time from the onset of 
convection up to the explosion; ${\rm\rho_{ign}}$ and ${\rm T_{ign}}$: central density  
and temperature at C-ignition; ${\rm\rho_{simm}}$ and ${\rm T_{simm}}$: 
central density and temperature at the onset of convection; M(\isotope{12}{C})$_{ini}$ 
and M(\isotope{12}{C})$_{fin}$:  amount of \isotope{12}{C} available in the accreted structure and 
that remaining at the explosion; $\Delta$M(\isotope{12}{C}): amount of 
\isotope{12}{C} consumed via nuclear burning up to the explosion;
${\rm M^{max}_{conv}}$: maximum extension of the convective zone; 
${\rm M_{exp}}$, $\rho_{\rm exp}$, $\eta_{\rm exp}$: mass coordinate where the 
explosion occurs and the corresponding values of  density and neutronization there; 
${\mathrm v_{exp}}$: maximum convective velocity at the explosion; 
${\overline\eta}_{\rm exp}$: neutronization at the explosion averaged over the convective zone; 
$\eta_{\rm c,0}$: initial value of central neutronization.}
\end{deluxetable}

As density increases, electron-captures on some nuclei become active. 
This affects the further evolution and, hence, the neutronization of the whole WD at the explosion 
in two different ways. In fact, electron-captures increase the local neutronization in the innermost zone 
of the accreting WD so that the final neutronization at the explosion is larger. Moreover, due to the neutrino 
emission associated with electron-capture and beta-decay, the local temperature is reduced. Hence, a larger amount of mass has to be 
accreted to attain C-ignition, which will occur at larger density and lower temperature. 
This occurrence affects the simmering phase, determining a larger final neutronization at the explosion, as 
illustrated in the lower panel of Fig.~\ref{fig1}, depicting the evolution of central $\eta$ of the ZSUN 
model (solid line). 
To disentangle the URCA processes effects, we compute two additional models, by excluding from the 
nuclear network some URCA processes. When the \isotope{32}{(S,P,Si)} and \isotope{56}{(Fe,Mn,Cr)} URCA triplets
are excluded (model NoSFe in Table~\ref{tab2} - dashed lines in Fig.~\ref{fig1}), the WD energy budget 
is only slightly altered, so that the physical conditions at the C-ignition and explosion epochs do not change appreciably. 
The final neutronization is $\Delta\eta\simeq 0.11\times 10^{-3}$ lower, corresponding to the 
\isotope{56}{Fe} and \isotope{32}{S} contribution already present in the initial WD. 
The exclusion of the \isotope{21}{(Ne,F)} URCA pair from the nuclear network (model NoNe21 in Table~\ref{tab2} 
- dotted lines in Fig.~\ref{fig1}) produces a small effect on the central neutronization before C-ignition
($\Delta\eta\simeq 4\times 10^{-6}$ - see lower panel in Fig.~\ref{fig1}), while the central temperature is about 
5\% larger. Hence, in model NoNe21 C-ignition occurs sooner and at lower density (see Table~\ref{tab2}) and
during the whole simmering phase up to the explosion the density profile remains lower than in the model ZSUN. This 
causes the neutronization at the explosion to be smaller by $\Delta\eta\simeq4\times 10^{-5}$, an order of 
magnitude larger than the difference before C-ignition. 

The neutron excess before C-ignition is affected also by electron-captures on several isotopes 
such as \isotope{41}{K}, which has an activation density of $\rho_{\text{ec}}=9.12\times 10^{8}$\gcc\, and 
mass fraction abundance $X_\odot ($\isotope{41}{K}$)=8.18\times 10^{-7}$. 
An interesting case is the electron-capture on \isotope{36}{Ar} (initial abundance $\mathrm{X_\odot ({^{36}Ar})=6.74\times  
10^{-5}}$ and $\rho_{\text{ec}}=1.71\times 10^{8}$\gcc), producing \isotope{36}{Cl} .
However, when the local density increases above $5.76\times 10^{8}$\gcc due to the continuous mass deposition, 
the $\mathrm{{^{36}Cl}(e^-,\nu_e){^{36}S}}$ process occurs.
\begin{figure}[t!]
   \includegraphics[width=\columnwidth]{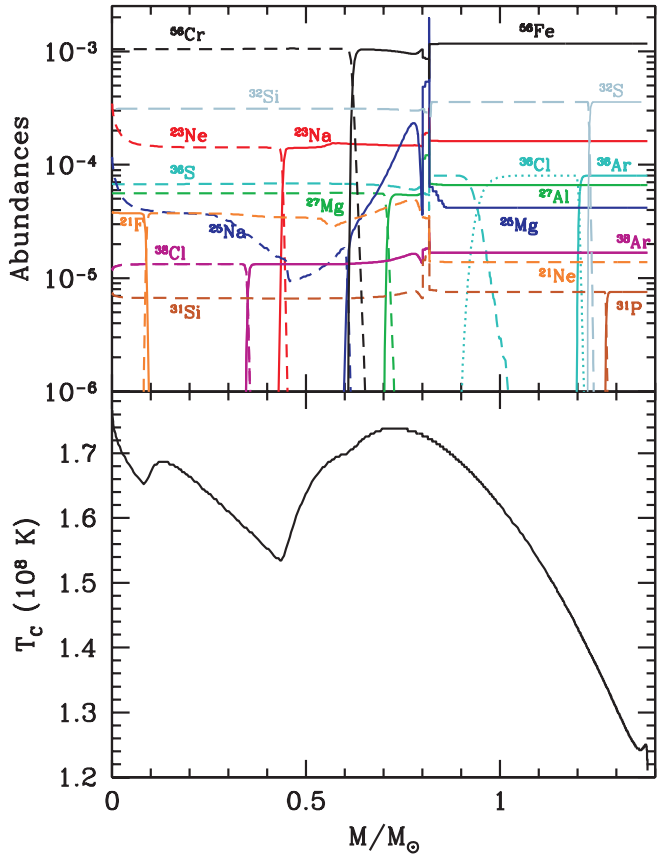}
    \caption{Thermal and chemical structure of the ZSUN model at the beginning of the simmering phase. 
                Isobars belonging to an URCA pair are identified with solid lines (electron-capturing one) and dashed lines (beta-decaying one). 
    \label{fig2}}
\end{figure}

Fig.~\ref{fig2} illustrates the interplay between heating, as determined by homologous compression, and cooling, 
due to URCA processes.
When central density exceeds $\rho_{\text{URCA}}$ for a given URCA pair, the center 
is cooled; as matter continues to be deposited, density increases along the whole accreting WD and  
the URCA shell moves outwards, cooling a zone progressively larger. On the contrary, due to compressional heating, 
the whole WD heats up. The interplay between these two processes determines the formation 
of minima in the temperature profile, as the one at M$\simeq$0.082 \msun, corresponding to the
\isotope{21}{(Ne,F)} URCA pair, or that at M$\simeq$0.430 \msun, corresponding to 
the \isotope{23}{(Na,Ne)} one. The earlier the URCA process activates 
during the accretion phase, the smaller the traces on the temperature profile at the C-ignition (e.g. the plateau at 
M$\simeq$0.620\msun, corresponding to the \isotope{25}{(Mg,Na)} URCA pair). The effects of 
URCA processes either ignited at the beginning of the accretion or with a small abundance of the 
\isotope{A}{Z} isobar are smeared off by the compressional heating.

Each jump in the neutronization displayed in the lower panel of Fig.~\ref{fig1} is the signature of 
a given URCA pair and/or double electron-capture triplet. Moreover, the boost in $\eta$ for $\rho\geq 4\times 10^{9}$ \gcc\, 
corresponds to the onset of C-burning, via $\mathrm{{^{12}C}({^{12}C},p){^{23}Na}}$, 
which revives the \isotope{23}{(Na,Ne)} URCA pair and activate the other major source of 
neutronization $\mathrm{{^{12}{C}}(p,\gamma){^{13}{N}}(e^-,\nu_e){^{13}C}}$.
For $\rho\geq5\times 10^9$ \gcc\, convection turns on, neutronized matter is diluted and neutronization increases 
at a lower rate. Finally, when local heating via C-burning largely dominates over the compressional heating, a large 
overpressure is determined and the whole structure starts to expand, while neutronization increases substantially.
\begin{figure}[t]
   \includegraphics[width=\columnwidth]{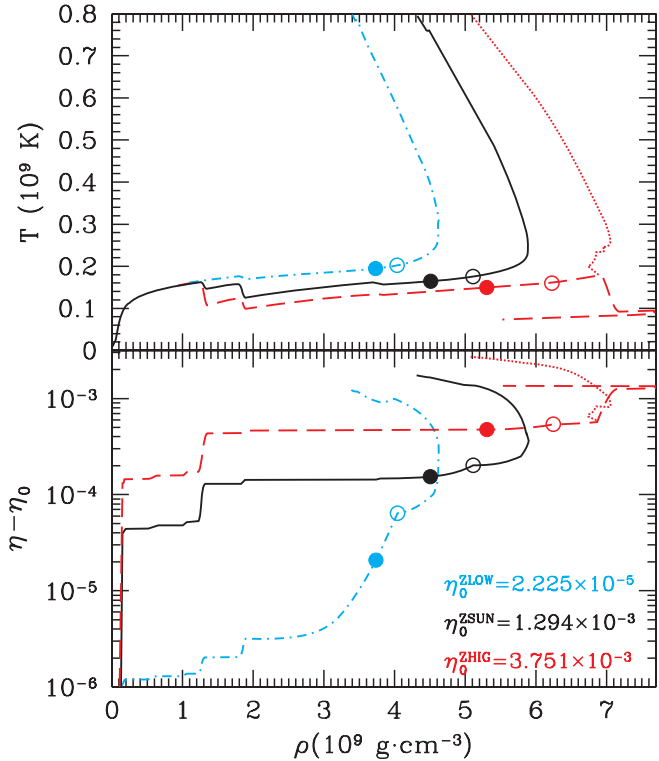}
    \caption{Evolution of the central temperature (upper panel) and neutron excess (lower panel) 
                as a function of the WD central density. Solid, dot-dashed and dashed lines refer to the ZSUN, ZLOW and ZHIG models, 
                respectively. Dotted lines refer to the mass coordinate where temperature is a maximum in the ZHIG model. 
                Filled and open dots mark the C-ignition and the simmering onset epochs, respectively.  
    \label{fig3}}
\end{figure}

The dependence of our results on the progenitor metallicity is summarized in 
Table~\ref{tab2} and illustrated in Fig.~\ref{fig3}. The ZLOW model is scarcely affected by 
the URCA shell induced cooling, due to the low initial abundance of isobars \isotope{Z}{A}. Hence, 
C-ignition occurs at lower density and higher temperature compared to the ZSUN model, causing  
a smaller increase of neutronization during the simmering phase.

In the ZHIG model, due to the larger initial abundance of isobars \isotope{A}{Z}, the cooling effects of URCA processes 
are larger compared to the ZSUN model and a larger amount of mass has to be 
accreted to trigger C-ignition at the center, occurring at very large density (see Table~\ref{tab2}). 
During the simmering phase the central density of the ZHIG model exceeds 
$7\times 10^9$\gcc, thus activating the $\mathrm{{^{20}Ne}(e^-,\nu_e){^{20}F}}$ reaction. Note that 
at this epoch, due to C-burning, \isotope{20}{Ne} abundance has increased from $\mathrm{X({^{20}Ne})_{ini}=3.18\times 10^{-3}}$ 
to  $\mathrm{X({^{20}Ne})=4.12\times 10^{-3}}$. Hence, due to efficient neutrino cooling, 
the innermost region of the WD becomes radiative while the location of maximum temperature shifts outwards
(dotted lines in Fig.~\ref{fig3}). 
The zones above the maximum temperature location decouples from the innermost region, 
which evolves at almost constant temperature. 
When thermal heating via nuclear burning overcomes the compressional heating, the whole structure expands 
and electron-captures on \isotope{20}{Ne} stop. Notwithstanding, the WD center can not heat up because 
thermal energy flows inwards from the burning region on a timescale definitively larger than the residual 
evolutionary time up to the explosion. As a matter of fact the explosion occurs off-center, 
at mass coordinate $\mathrm{M_{exp}\simeq 0.0149}$\msun. 

\section{Conclusions}\label{s:conclu}
We have computed the evolution of three accreting WDs from the beginning of the the mass transfer process up to the 
thermal runaway, through the simmering phase, paying special attention to the treatment of convection 
and URCA processes. We find that weak processes play a crucial role in determining the physical conditions and 
neutronization at the explosion. 
Contrary to the claims by \citet{piro2008,chamulak2008,martinez2016}, we find that the simmering phase 
acts as an amplifier of the initial metallicity differences, determining quite different neutron excesses, densities 
and convective velocities at the explosion. 
The comparison of our models with those present in the extant literature  
is far from being straightforward, due to the different input physics adopted in the various numerical 
simulations (e.g., initial WD models, nuclear network, equation of state, nuclear reaction rates, 
mixing algorithm, etc.), so we will address this issue in a forthcoming paper.

In principle, one could define a final-to-initial neutron-excess relation (FINE), that could be used to decode the neutron 
excesses measured in SNIa explosions or remnants and, then, to deduce the progenitor metallicity. 
By linearly interpolating the values listed in Tab. \ref{tab2} of $\mathrm{\bar{\eta}_{exp}}$ as a function of Z 
for models ZLOW, ZSUN and ZHIG, we obtain
\begin{eqnarray}\label{eq1}
\overline{\eta}_{exp}=1.285 \eta_0 & + & 6.649\times 10^{-4},\nonumber\\ 
R^2 & = &1.00\, .
\end{eqnarray}
By recalling that $\eta_0$ depends only on the initial metallicity,
\begin{equation}\label{eq1c}
\eta_0=9.380\times 10^{-2} Z - 4.656\times 10^{-7} \,,
\end{equation}
Eq.~\ref{eq1} can be written as:
\begin{equation}\label{eq2}
\overline{\eta}_{exp}=1.205\times 10^{-1} Z + 6.643\times 10^{-4}\,,
\end{equation}
and the latter could be inverted to obtain:
\begin{equation}\label{neweq}
Z=8.295	\overline{\eta}_{exp} -5.507\times 10^{-3}\,.
\end{equation}
Note that the coefficients in all the previous equations depend on the assumed distribution of metals 
in the CO WD progenitors.

\citet{badenes2008,park2013} measured the Mn and Cr mass for the Tycho and Kepler remnants and 
derived $\overline{\eta}_\mathrm{exp,Tycho}=4.36\times10^{-3}$ and $\overline{\eta}_\mathrm{exp,Kepler}=4.55\times10^{-3}$. 
By assuming that these values represent the neutronization at the explosion epoch and that the latter is equal to 
that of the initial WD, mainly determined by the \isotope{22}Ne abundance ($\mathrm{\eta_0 = Z/11}$), 
they derived the progenitor metallicity of Tycho's and Kepler's supernova to be 
$Z_\mathrm{Tycho}=0.048=3.5Z_\odot$ and $Z_\mathrm{Kepler}=0.050=3.6Z_\odot$, respectively.
By using Eq.~\ref{neweq}, we derive $Z_\mathrm{Tycho}=0.0307=2.22Z_\odot$ and $Z_\mathrm{Kepler}=0.0322=2.34Z_\odot$,
in fairly good agreement with the evidence that stars in the thin disk of the Galaxy have metallicity lower than $3Z_\odot$. 

According to our results, the higher the metallicity, the larger $\mathrm{\overline{\eta}_{exp}}$ 
and $\rho_{exp}$ and, hence, the lower the amount of \isotope{56}{Ni} synthesized during the explosion.
Hence, low metallicity progenitors are expected to produce brighter SNe Ia. The explosion phase
of models discussed here will be presented in a forthcoming paper.

We remark that our results do not apply to SNe Ia progenitor models involving WDs 
with total mass definitively lower than the Chandrasekar limit.

The variation of $\overline{\eta}$ during the accretion and simmering phases depends mainly on the 
heavy elements abundance in the WD progenitor. However, the physical properties during the accretion phase 
depends also on \mdot\, and to lesser extent on the WD initial mass. 
In particular, for a fixed \mwd, the lower \mdot, the larger ${\mathrm M_{acc}}$ and the higher $\mathrm{\rho_{ign}}$ 
\citep[e.g. see][]{piersanti2003a}. This should affect the evolution during the simmering phase, producing 
a spread in $\mathrm{\overline{\eta}_{exp}}$ and $\mathrm{\rho_{exp}}$.

In this work we ignored the effect of gravitational settling of metals during the WD cooling phase. 
Since we found that their initial presence affects the evolution of the accreting WD, 
we plan to investigate the impact of sedimentation on $\overline{\eta}$ during accretion and simmering phases.

\acknowledgments
L.P., S.C. and O.S. acknowledge founding from the PRIN-MIUR grant 20128PCN59; 
E.B. and I.D acknowledge founding from the MINECO-FEDER grant AYA2015-63588-P; 
G.M.P. is partly supported by the Deutsche Forschungsgemeinschaft through contract SFB 1245.

\bibliographystyle{aasjournal} 
\bibliography{piersanti} 

\end{document}